\documentclass[11pt,subeqn,fleqn]{article}
\usepackage{graphicx}
\usepackage{latexsym}
\usepackage{amsfonts}
\usepackage{graphicx}
\usepackage{epic}
\usepackage{graphicx,lscape,fancyhdr,array}
\usepackage{amsthm,latexsym,multibox}
\usepackage{amssymb,amsfonts,amsbsy}

\newcommand{\ft}{\footnotesize}
\def\a{\alpha}
\def\b{\beta}

\def\d{\delta}

\def\l{\lambda}
\def\G{\Gamma}

\def\m{\mu}

\def\o{\omega}

\def\lag{{\mathfrak{g}}}
\def\lah{{\mathfrak{h}}}
\def\lan{{\mathfrak{n}}}
\def\las{{\mathfrak{s}}}
\def\lap{{\mathfrak{p}}}
\def\lak{{\mathfrak{k}}}
\def\cO{{\cal O}}
\def\cV{{\cal V}}
\def\be{\begin{equation}}
\def\ee{\end{equation}}
\def\beq{\begin{eqnarray}}
\def\eeq{\end{eqnarray}}
\def\no{\nonumber}
\newsavebox{\uuunit}
\sbox{\uuunit}
    {\setlength{\unitlength}{0.825em}
     \begin{picture}(0.6,0.7)
        \thinlines
        \put(0,0){\line(1,0){0.5}}
        \put(0.15,0){\line(0,1){0.7}}
        \put(0.35,0){\line(0,1){0.8}}
       \multiput(0.3,0.8)(-0.04,-0.02){12}{\rule{0.5pt}{0.5pt}}
     \end {picture}}

\begin{document}

\begin{titlepage}

\begin{flushright}
ULB-TH/04-23\\
AEI-2004-078\\
\end{flushright}
\vskip 1.0cm

\begin{centering}
\mbox{}\\[10mm]
\begin{center}{\LARGE \bf  Higher spin fields from indefinite\\
    Kac--Moody algebras}\\[18mm]
{\bf Sophie de Buyl}\\
{\sl Physique Th\'eorique et Math\'ematique and
International Solvay Institutes}\\
{\sl Universit\'e Libre de Bruxelles}\\
{\sl C.P. 231, B-1050, Bruxelles, Belgium}\\[5mm]
{\bf Axel Kleinschmidt}\\
{\sl Max-Planck-Institut f\"ur Gravitationsphysik}\\
{\sl M\"uhlenberg 1, D-14476 Golm, Germany}\\
axel.kleinschmidt@aei.mpg.de
\end{center}

\vspace{10mm}

\end{centering}
\renewcommand{\abstract}{\begin{center}\bf
    Abstract\\[3mm]\end{center}}
\begin{abstract}
The emergence of higher spin fields in the Kac--Moody theoretic
approach to M-theory is studied. This is based on work done by
Schnakenburg, West and the second author. We then study the relation
of higher spin fields in this approach to other results in different
constructions of higher spin field dynamics. Of particular interest
is the construction of space-time in the present set-up and we
comment on the various existing proposals.
\\ \vskip 24pt
\begin{center}
{\it Based on a talk presented by A. Kleinschmidt at the
First Solvay Workshop on Higher-Spin Gauge Theories held in
Brussels on May 12--14, 2004}
\end{center}
\vskip 5pt

\end{abstract}

\vfill
\end{titlepage}

\section{Introduction and Motivation} \
\indent M-theory is usually thought to comprise the different
string theories and is therefore related to higher spin (HS) field
theories in the
sense that string theories contain HS. Indeed, the perturbative spectra
of the different string theories characteristically contain
massive HS fields if the tension of the string
is finite and massless HS fields in the limit of vanishing
tension. The tensionless
limit $\alpha'\rightarrow\infty$, in which all perturbative states
become massless, is of particular interest because of the restoration
of very large
symmetries \cite{Gross:1988ue}. One possible approach to
M-theory is based on such infinite-dimensional symmetries; work along
these lines can be found in
\cite{Cremmer:1997ct,West:2001as,BJall1,Damour:2002cu,Englert:2003py}
and references therein. Our focus here will be the Kac--Moody theoretical
approach of \cite{West:2001as} and
\cite{Damour:2002cu,Englert:2003py}.
The aim of the present discussion is to show
how higher spin fields arise naturally as objects in this algebraic
formulation and speculate how a dynamical scheme for these fields
might emerge. Before studying the details of the HS fields, let us briefly
review the origin of the Kac--Moody algebras we are going to consider.
\newline

\indent It is a longstanding conjecture that the reduction of certain
(super-)gravity systems to very low space-time dimensions exhibits
infinite-dimensional symmetry algebras \cite{Ju82}. As these sectors
are normally thought of as low energy limits of M-theory, these
conjectures extend to M-theory, albeit with certain modifications
\cite{Hull:1994ys}. The precise statement for the (super-)gravity
systems is that the scalar sector (with dualisation of fields to
scalar fields whenever possible)
is described by a non-linear sigma model $G/K(G)$ where $G$ is
some group and $K(G)$ its maximal compact subgroup \cite{Ju82}.\newline

\noindent As an example, the chain of these so-called `hidden
symmetries' in the case of $D_{max}=11$, $N=1$ supergravity is
displayed in the table below. The scalar sector
of this theory reduced to $D$ space-time dimensions is described by a
scalar coset model $G/K(G)$ ($G$ is in the
split form)  determined by

\begin{center}
\begin{tabular}{|c|c|c|}
\hline  $D$ & $G$ & $K$ \\
\hline  $9$ & $SL(2,R) \times SO(1,1)$ & $SO(2)$  \\
\hline $8$ &  $S(3,R) \times SL(2,R)$  & $U(2)$  \\
\hline $7$ &$SL(5,R)$  & $USp(4)$ \\
\hline $6$& $SO(5,5)$  & $USp(4) \times USp(4)$  \\
\hline $5$ & $E_{6(6)}$ &  $USp(8)$ \\
 \hline $4$&
$E_{7(7)}$& $SU(8)$ \\
\hline $3$&
$E_{8(8)}$& $Spin(16) / Z_2$\\
\hline  \hline $2$&
$E_{9(9)}$ & $K(E_9)$ \\
\hline $1$ &$E_{10(10)}$ & $K(E_{10})$ \\
\hline $(0$ &$E_{11(11)}$ & $K(E_{11}))$ \\
\hline
\end{tabular}
\end{center}
After the reduction to three dimensions one obtains Kac--Moody
theoretic extensions of the exceptional $E$-series. The two-dimensional
symmetry $E_{9}$ is an affine symmetry \cite{Ju82,BrMa87}, below two
dimensions the expected symmetry is the hyperbolic extension
$E_{10}$ and in one dimension one formally finds the Lorentzian
algebra $E_{11}$. Three dimensions are special since there all
physical (bosonic) degrees of freedom can be converted into
scalars.
The  procedure inverse to dimensional reduction, called
oxidation, of a theory has also been studied
\cite{Ju81a,Cremmer:1999du,Ke03a}. There one starts with a scalar
coset model $G/K$ in three space-time dimensions for each simple $G$
and asks for a higher-dimensional theory whose
reduction yields this hidden symmetry. The answer is known for all
semi-simple $G$ and was presented in \cite{Cremmer:1999du} and will be
recovered in table \ref{fcres} below. For conciseness of notation we
denote the oxidized theory in the maximal dimension by $\cO_G$. In
this language,
maximal eleven-dimensional supergravity is denoted by $\cO_{E_8}$.
\newline

\indent A different motivation for studying infinite-dimensional
Kac--Moody algebras comes from cosmological billiards. Indeed,
it has been shown recently that the dynamics of the gravitational
scale factors becomes equivalent, in the vicinity of a spacelike
singularity, to that of a relativistic particle moving freely on
an hyperbolic billiard and bouncing off its walls
\cite{Belinsky:1970ew, Belinsky:1982pk, Damour:2002et,Damour:2002fz}. A
criterion for the gravitational dynamics to be chaotic is that the
billiard has a finite volume \cite{Damour:2001sa}. This in turn
stems from the remarkable property that the billiard walls can be
identified with the walls of the fundamental Weyl chamber of a
hyperbolic Kac--Moody algebra.\footnote{Recall that an algebra is
  hyperbolic if  upon the deletion of any node of the Dynkin diagram
  the remaining  algebra is a direct sum of  simple or affine Lie
  algebras.} Building on  these observations it has been shown that a
null geodesic motion of a relativistic particle in the coset space
$E_{10}/K(E_{10})$ can be mapped to the bosonic dynamics of $D=11$
supergravity reduced to one time dimension
\cite{Damour:2002cu,Damour:2002et}.
\newline

\indent It is at the heart of the proposal of \cite{West:2001as}
that the hidden symmetries of the reduced theory are already
present in the unreduced theory. Furthermore, the symmetry groups
actually get extended from the finite-dimensional $G$ to the
Lorentzian triple extension $G^{+++}$ (also called
very-extension).\footnote{For the process of very-extension see
also \cite{Gaberdiel:2002db,En,Kleinschmidt:2003mf}.}
To every finite-dimensional $G$ we wish to associate a model
possessing a non-linearly realized $G^{+++}$ symmetry which might be
called the M-theory corresponding to $G$, and denoted by $\cV_G$.
To try to construct M-theory in the sense given above, the  basic
tool is a non-linear sigma model based on Kac--Moody algebras in
the maximal number of dimensions (corresponding to the oxidized
$\cO_G$). Basic features about non-linear sigma models associated
with the coset $G/ K(G)$ are recalled in appendix A. M-theories
constructed in this way possess an infinite-dimensional symmetry
structure and contain infinitely many fields. Some of the infinitely
many fields may be  auxiliary, however.\footnote{In order to determine
  which fields are auxiliary one would require a (still missing)  full
  dynamical understanding of the   theory.}  In addition one can hope
to relate some of the infinitely many other fields to the perturbative
string spectrum. Since the underlying Kac--Moody algebra is not
well-understood the analysis is complicated and incomplete,
and this identification of states has not
been carried out. Below we will present the result for the
supergravity fields. In addition, the precise dynamical theory for
all fields is mostly elusive to date. There remain a number of
important challenges in the full understanding of these
M-theories, which will be highlighted at the relevant points.
\newline

\indent Since $E_{11}$ will be our guiding example, we briefly review
the evidence for the conjecture that M-theory possesses an $E_{11}$ symmetry.
\newline

\begin{itemize}
\item It has been shown that the bosonic sector of $D=11$
supergravity can be formulated as the simultaneous non-linear
realization of two finite-dimensional Lie algebras
\cite{West:2000ga}. The two corresponding groups,
whose closure is taken, are
the eleven-dimensional conformal group and a group called
$G_{11}$ in \cite{West:2000ga}. The generators of $G_{11}$ contain
the generators $P_a$ and $K^a{}_b$ of the
group of affine coordinate transformations $IGL(11)$ in eleven
dimensions and the closure with the conformal group
will therefore generate infinitesimal general
coordinate transformations \cite{Og73}. The precise structure of the
Lie algebra of $G_{11}$ is given by

\beq \ [K^a{}_b,K^c{}_d]&=&\delta _b^c K^a{}_d - \delta _d^a
K^c{}_b \nonumber \\ \ [K^a{}_b,P_c]&=& - \delta _c^a P_b
\nonumber
\\ \ [K^a{}_b, R^{c_1\ldots c_6}]&=& -6 \delta _b^{[c_1}R^{c_2\ldots
c_6]a}, \nonumber \\
 \ [K^a{}_b, R^{c_1\ldots c_3}]&=& 3\delta _b^{[c_1}R^{c_2
 c_3]a}, \nonumber \\
 \ [ R^{c_1\ldots c_3}, R^{c_4\ldots c_6}]&=& 2 R^{c_1\ldots c_6}.
\eeq

The additional totally anti-symmetric
generators $R^{c_1\ldots c_3}$ and $R^{c_1\ldots
c_6}$ are obviously related to the three-form gauge potential of
$N=1$, $D=11$ supergravity and its (magnetic) dual.
To write the equations of motion of the bosonic sector of
$D=11$ supergravity one should first consider the following coset
element of $G_{11}$ over the Lorentz group $H_{11}$,

\beq
v(x) = e^{x^a P_a} e ^{h_{a}{}^b K^a{}_b} \exp
{({A_{c_1\ldots c_3} R^{c_1\ldots c_3}\over 3!}+ {A_{c_1\ldots
c_6} R^{c_1\ldots c_6}\over 6!})}\in G_{11}/H_{11}
\eeq

\noindent The fields  $h^a{}_b$, $A_{c_1\ldots c_3}$ and
$A_{c_1\ldots c_6}$ depend on the space-time coordinate
$x^\mu$.  Notice that the
non-linear sigma model $G_{11} / H_{11}$ does not produce the
bosonic equations of motion of $D=11$ supergravity. The Cartan
forms $\partial_\mu v v^{-1}$ constructed from this coset element
for instance do not give rise to the antisymmetrized field strengths.
Only after the
non-linear realization with respect to the coset of the
conformal group with the {\em same} Lorentz group is also
taken does one
obtain the correct field strengths and curvature terms of the
supergravity theory \cite{West:2000ga}. Then it is natural to
write down the equations of motion in terms of these fields,
which then are the full non-linear field equations. Similar
calculations were done for the type ${\rm IIA}$, ${\rm IIB}$ and
massive ${\rm IIA}$ supergravity \cite{West:2000ga,SchnWe01,SchnWe02},
and for the closed bosonic
string \cite{LaWe01}.

\item It has been conjectured that an extension of this theory
has the rank eleven Kac--Moody algebra called $E_{11}$ as a
symmetry \cite{West:2001as}. $E_{11}$ is the very-extension of
$E_8$ and is therefore also sometimes called $E_8^{+++}$.
This extension is suggested because,
if one drops the momentum generators $P_a$, the coset
$G_{11}/H_{11}$ is a
truncation of the coset $E_{11}/K(E_{11})$, as we will see below.
Note that dropping the translation generators $P_a$ in a sense
corresponds to forgetting about space-time since the
field conjugate to $P_a$ are the space-time coordinates $x^a$.
The translation operators $P_a$ will be re-introduced in section 4
where we will also discuss the r\^ole of space-time in more detail.

\item The algebraic structure encoded in $E_{11}$ has been shown
to control transformations of Kasner solutions \cite{En} and
intersection rules of extremal brane solutions \cite{EnHoWe03}.
Furthermore, the relation between brane tensions in IIA and IIB
string theory (including the D8 and D9 brane) can be deduced
\cite{West:2004st,West:2004kb}.
\end{itemize}

\noindent As mentioned above, similar constructions
can be done for all very-extended $G^{+++}$. In
the sequel, the attention is focussed on these models.
\newline

The paper is organized as follow. Chapter 2 is devoted to studying
the field content of the Kac--Moody models $\cV_G$. A brief
introduction to Kac--Moody algebras is given in section 2.1. A
level decomposition of the generators of a Kac-Moody algebra with
respect to a $\mathfrak{gl}(D)$ subalgebra is explained in section
2.2. Such decompositions are of interest since the
$\mathfrak{gl}(D)$ subalgebra represents the gravitational degrees
of freedom; the other generators are in tensorial representations
of $\mathfrak{gl}(D)$ and mostly provide higher spin fields. In
section 2.3, the field content of the model $\cV_G$ is given for
all $G^{+++}$ and the (super-)gravity fields of $\cO_G$ are
identified. Another remarkable feature of the algebras is how
their Dynkin diagrams support different interpretations, in
particular T-dualities. This is exemplified in section 2.4. In
chapter 3, we offer a few remarks on the relation of the KM
fields to higher spin theories. The emergence of space-time
in the context of very-extended algebras is envisaged in chapter
4, where we also explain some diagrammatic tricks.
The conclusions are given in the last chapter.

\setcounter{equation}{0}

\section{Field content of Kac--Moody algebras}

\subsection{Brief definition of Kac--Moody algebra}

\indent Let us define a Kac--Moody algebra $\lag$  via its
Dynkin diagram with $n$ nodes and links between these nodes. The
algebra is a Lie algebra with Chevalley generators $h_i, \ e_i, \
f_i$ ($i=1,\ldots,n$) obeying the following relations \cite{Kac}

\begin{eqnarray}
\ [ e_i ,f_j ] &=& \d_{ij} h_i \nonumber \\
\ [h_i ,e_j] &=& A_{ij} e_j \nonumber \\
\ [h_i ,f_j] &=& -A_{ij} f_j \nonumber \\
\ [h_i ,h_j] &=& 0 \nonumber
\end{eqnarray}

\noindent where $A_{ii}=2$ and $-A_{ij}$ ($i \neq j$) is a
non-negative integer related to the number of links between the
$i^{\mathrm{th}}$ and $j^{\mathrm{th}}$ nodes. The so-called Cartan
matrix $A$ in addition satisfies $A_{ij}=0 \Leftrightarrow A_{ji}=0$.
The generators must
also obey the Serre relations,

\begin{eqnarray}
(\mathrm{ad}\,e_i)^{1-A_{ij}} e_j &=& 0 \nonumber \\
(\mathrm{ad}\,f_i)^{1-A_{ij}} f_j &=& 0 \nonumber
\end{eqnarray}
A root $\a$ of the algebra is a non-zero linear form on the Cartan
subalgebra $\lah$ (= the subalgebra generated by the $\{ h_i \ \arrowvert
\  i = 1,... ,n \}$)\footnote{Here we assume the Cartan matrix $A$ to
  be non-degenerate.} such that
$$\lag_{\a} = \{ x \in \lag \ \arrowvert [ h, x ] =
\a (h) x \ \forall h \in \lah \}$$ is not empty. $\lag$ can be
decomposed in the following triangular form,

$$ \lag = \lan_- \oplus \lah \oplus \lan_+   $$ or according to the root spaces,
$$ \lag = \oplus_{\a \in \Delta} \lag_{\a} \oplus \lah$$ where
$\lan_-$ is the direct sum of the negative roots spaces, $\lan_+$ of the
positive ones and $\lah$ is the Cartan subalgebra. The dimension of
$\lag_{\a}$ is called the multiplicity of $\a$. These
multiplicities obey the Weyl--Kac character formula

$$ \Pi_{\a \in \Delta_+} (1 - e^{\a})^{\mathrm{mult}\a} = \sum_{w
\in W}  \epsilon(w) e^{w(\rho)-\rho} $$

\noindent The sum is over the Weyl group which in the case of
interest here is infinite. $\epsilon(w)$ is the parity of $w$ and
$\rho$ is the Weyl vector. This formula cannot be solved in closed
form in general. We will normally denote by $G$ the (simply-connected,
formal) group associated to $\lag$.\newline

Our interest is focussed here on a class of Kac--Moody algebras
called \textit{very-extensions} of simple Lie algebras. They are
the natural extension of the over-extended algebras which are
themselves extensions of the affine algebras. The procedure for
constructing an affine Lie algebra $\lag^{+}$ from a simple one $\lag$
consist in the addition of a node to the Dynkin diagram in a
certain way which is related to the properties of the highest root
of $\lag$. One may also further increase by one the rank of the
algebra $\lag^{+}$ by adding to the Dynkin diagram a further node
that is attached to the affine node by a single line. The
resulting algebra $\lag^{++}$ is called the over-extension of $\lag$. The
very-extension, denoted $\lag^{+++}$, is found by adding yet another
node to the Dynkin diagram that is attached to the over-extended
node by one line \cite{Gaberdiel:2002db}.

A further important concept in this context is that of the {\em compact
form $K(\lag)$} of $\lag$ \cite{Kac}, which we here define as the fixed point
set under the compact involution $\o$ mapping
$$
\o(e_i)=-f_i,\qquad\o(f_i)=-e_i,\qquad\o(h_i)=-h_i.
$$
An element in the coset space of the formal groups $G/K(G)$ then
can be parametrized as
\beq\label{cosel}
v=\exp(\sum_i \phi_i h_i) \exp(\sum_{\a>0} A_\a E_\a),
\eeq
where the (infinitely many)
positive step operators $E_\a$ can have multiplicities greater than one
for imaginary roots $\a$.

\subsection{Decomposition of a KM algebra under the
action of a regular subalgebra}

In order to construct a non-linear sigma model associated with a
Kac--Moody algebra, e.g. $E_{8}^{+++}$, one needs to consider
infinitely many step operators $E_{\alpha}$ and therefore
infinitely many corresponding fields $A_{\a}$
because the algebra is infinite-dimensional.\footnote{See also appendix
A for the notation.} To date, only truncations of this model
can be constructed. A convenient
organization of the data is to decompose the set of generators of
a given Kac--Moody algebra $\lag$ with respect to a finite regular
subalgebra $\las$ and a corresponding {\em level
decomposition}.\footnote{We restrict to finite-dimensional
subalgebras $\las$ in order to have only a finite number of
elements for each level \cite{Damour:2002cu,NiFi03}.}
The `level' provides a gradation on $\lag$ and the assignment is
as follows: generators in $\lah$ are at level $\ell=0$ and
elements in $\lag_{\a}$ derive their level from the root $\a$,
usually as a subset of the root labels.
The following example is taken to illustrate
the notion of level.
\begin{quote} \textit{Level of a root}: Consider
the algebra $\lag=E_6^{+++}$ with simple roots $\a_i$
($i=1,\ldots,9$) labelled according to the Dynkin diagram:
\vspace{.3cm}
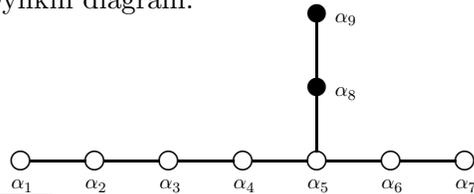
\begin{center}
\scalebox{.7}{
\begin{picture}(180,60)
\put(5,-5){$\alpha_{1}$} \put(45,-5){$\alpha_2$}
\put(85,-5){$\alpha_3$}
 \put(125,-5){$\alpha_4$}
  \put(165,-5){$\alpha_5$} \put(205,-5){$\alpha_6$}
  \put(245,-5){$\alpha_7$}
  \put(180,45){$\alpha_8$}   \put(180,85){$\alpha_9$}
\thicklines \multiput(10,10)(40,0){7}{\circle{10}}
\multiput(15,10)(40,0){6}{\line(1,0){30}}
\put(170,50){\circle*{10}} \put(170,15){\line(0,1){30}}
\put(170,90){\circle*{10}} \put(170,55){\line(0,1){30}}
\end{picture}
} \end{center} We want to make a level decomposition of $E_6^{+++}$
under its $\las=A_7$ subalgebra corresponding to the sub-Dynkin
diagram with  nodes from 1 to 7. This singles out the nodes $8$ and
$9$ which do not belong to the $A_7$ subdiagram. Any positive root
$\a$ of $E_6^{+++}$ can be written as $\a = \sum_{i=1}^9 m_i \a_i =
\sum_{s=1}^7 m_s \a_s+ \sum_{g= 8,9} \ell_g \a_g$ with $m_s$ and
$\ell_g$ non-negative integers. Here, $\ell_8$ and $\ell_9$ are
called respectively the $\a_8$  level and the $\a_9$ level of
$\a$.
\end{quote} The truncation consists of considering only the generators
of the lowest levels; the coset element (\ref{cosel}) is
then calculated by using only the fields corresponding to these
generators. The question of which cut-off to take for the truncation
will be answered universally below.\newline

\noindent The adjoint action of $\las$ on $\lag$
preserves the level, therefore the space of
generators on a given level is a (finite-dimensional)
representation space for a representation of $\las$ and hence
completely reducible. The
fields $A_{\a}$ associated with roots of a given level $\{\ell_g\}$
will be written as representations of $\las$. \newline

\noindent In order to obtain an interpretation of the fields
associated with these generators as space-time fields, it is
convenient to choose the regular subalgebra to be
$\las=\mathfrak{sl}(D)$ for some $D$; these are always enhanced to
$\mathfrak{gl}(D)$ by Cartan subalgebra generators associated with
the nodes which were singled out. The reason for the choice of an
$A$-type subalgebra is that one then knows how the resulting
tensors transform under $\mathfrak{so}(D)=K(\las)\subset K(\lag)$.
In the spirit of the non-linear realization explained in the
introduction, this $\mathfrak{so}(D)$ plays the r\^ole of the
local Lorentz group.\footnote{The real form $\mathfrak{so}(D)$ is
actually not the correct one for an interpretation as Lorentz
group. In order to obtain the correct $\mathfrak{so}(D-1,1)$ one
needs to modify the compact involution to a so-called `temporal
involution' \cite{Englert:2003py}. As discussed in
\cite{Keurentjes:2004bv,Keurentjes:2004xx} this leads to an
ambiguity in the signature of space-time since Weyl-equivalent
choices of involution result in inequivalent space-time signatures
on the `compact' part of the subalgebra $\las$.
We will not be concerned with this
important subtlety here as it does not affect the higher spin
field content.}
In particular, the level $\ell=0$ sector will give rise to the
coset space $GL(D) / SO(D)$ since it contains the adjoint
of the regular $\mathfrak{gl}(D)$ subalgebra. This
is the right coset for the
gravitational vielbein which is an invertible matrix defined up to a
Lorentz transformation.\footnote{For a recent
`stringy' decomposition of $E_{10}$ under a subalgebra of type $D$ see
\cite{KlNi04}.} If the rank of $\lag$ is $r$ then
there will also be $r-D$ additional dilatonic scalars at level
$\ell=0$ from the remaining Cartan subalgebra generators.
\newline

\noindent Let us study in more detail which representations
of $\las$ can occur at a given level. The necessary
condition for a positive root $\a  = \sum_s m_s \a_s +
\sum_g \ell_g \a_{g}$ of $\lag$ to generate a lowest weight
representation of $\las$ are that

$$ p_s = - \sum_t A_{st} m_t - \sum_g A_{sg} \ell_g \ge 0$$
where $A$ is the Cartan matrix of $\lag$,  the $\{p_s\}$ are
the Dynkin labels of the corresponding lowest weight
representation.\footnote{Actually the lowest weight then
has Dynkin labels $\{-p_s\}$ since we have introduced an additional
minus sign in the conversion between the different labels which is
convenient for the class of algebras we are considering here.}
We can also rephrase this condition by realizing that a
representation with labels $\{p_s\}$ at level $\{\ell_g\}$
corresponds to a root with coefficients

$$ m_s = - \sum_t (A^{-1}_{\mathrm{sub}})_{st} p_t  \sum_{t,g}-
( A^{-1}_{\mathrm{sub}}) _{st} A_{tg} l_g \geq 0 $$
and these are non-negative integers if they are to belong to a
root of $\lag$. Here,
$A_{\mathrm{sub}}$ is the Cartan matrix of $\las$. Besides this
necessary condition one has to check that there are elements
in the root space of $\a$ that can serve as highest weight
vectors. This requires calculating the multiplicity of $\a$ as a
root of $\lag$ and its weight multiplicity in other (lower)
representations on the same level. The number of independent
highest weight vectors is called the {\em outer multiplicity} of
the representation with labels $\{p_s\}$ and usually denoted by
$\m$.
\newline

\begin{quote}
\textit{Example}:
$\lag = E_8^{+++}$, which is the very extension of $E_8$,
can be decomposed  w.r.t its
subalgebra $ \las = A_{10}=\mathfrak{sl}(11)$ associated with the sub-Dynkin
diagram with nodes from 1 to 10,

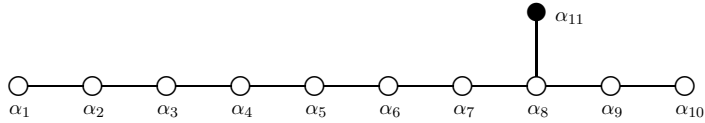
\begin{center} \scalebox{.7}{
\begin{picture}(180,60)
\put(-35,-5){$\a_{1}$} \put(5,-5){$\alpha_{2}$}
\put(45,-5){$\alpha_3$} \put(85,-5){$\alpha_4$}
 \put(125,-5){$\alpha_5$}
  \put(165,-5){$\alpha_6$} \put(205,-5){$\alpha_7$}
  \put(245,-5){$\alpha_8$}   \put(285,-5){$\alpha_9$}
  \put(325,-5){$\alpha_{10}$}
  \put(260,45){$\alpha_{11}$}
\thicklines \multiput(-30,10)(40,0){10}{\circle{10}}
\multiput(-25,10)(40,0){9}{\line(1,0){30}}
\put(250,50){\circle*{10}} \put(250,15){\line(0,1){30}}

\end{picture}
} \end{center} One finds that the level $\ell\equiv\ell_{11} = $
0, 1, 2 and 3 generators are in a lowest weight representation of
$A_{11}$ with Dynkin labels given in the following table,
\begin{center}
\begin{tabular}{|c|c| l|}
\hline

$\ell$ & Dynkin labels & Tensor\\
\hline

0 & $[1, \ 0, \ 0, \ 0, \ 0 , \ 0 , \ 0 , \ 0 , \ 0, \ 1] + [0, \
0, \ 0, \ 0, \ 0 , \ 0 , \ 0 , \ 0 , \ 0, \ 0]$ & $K_a^{\ b}$ \\
1 & $[0, \ 0, \ 0, \ 0, \ 0 , \ 0 , \ 0 , \ 1 , \ 0, \ 0]$ &
$R^{a_1 a_2 a_3}$ \\
2 & $[0, \ 0, \ 0, \ 0, \ 1 , \ 0 , \ 0 , \ 0 , \ 0, \ 0]$ &
$R^{a_1 a_2 a_3 a_4 a_5 a_6}$ \\
3 & $[0, \ 0, \ 1, \ 0, \ 0 , \ 0 , \ 0 , \ 0 , \ 0, \ 1]$ &
$R^{a_1 ... a_8,b}$ \\
... & & \\
\hline
\end{tabular}
\end{center} The level zero generators are associated with the
gravitational degrees of freedom as explained above. The level one
field can be recognized as the three form potential of the
supergravity in $D=11$ and the level two field correspond to its
dual. Notice that up to level 2 the generators are just the fields
of $G_{11}$, mentioned in the introduction, except for the
momentum generators $P_a$.
The level three tensor corresponds to
a Young tableau of the form
\begin{center}
\scalebox{.9}{\begin{picture}(30,30)(0,-10)
\multiframe(0,10)(10.5,0){1}(10,10){\ft$a_1$}
\multiframe(10.5,10)(10.5,0){1}(10,10){\ft$b$}
\multiframe(0,-0.5)(10.5,0){1}(10,10){\ft$a_2$}
\multiframe(0,-18)(10.5,0){1}(10,17){$ $} \put(4,-13.5){$\vdots$}
\multiframe(0,-28.5)(10.5,0){1}(10,10){\ft$a_8$}
\end{picture}}
\end{center}
$\ $ \newline \noindent and is associated with the dual of the
graviton
\cite{Hull:2001iu,Curtright:1980yk,Bekaert:2002uh}.\footnote{There
is
  a subtlety here since the $\mathfrak{gl}(11)$ tensor associated
  with this mixed symmetry vanishes after antisymmetrization over all
  indices. This means that one cannot accomodate the trace of the spin
  connection in this dual picture.}
The dualization of Einstein's field equations only works
at the linearized level however
\cite{Bekaert:2002uh}. One might speculate that a dualization of
the non-linear equations will probably require some of
the remaining fields in the infinite list of tensors above level 3.
In summary, the lowest level generators ($\ell=0,1,2,3$) of
the $E_8^{+++}$ algebra correspond to the degrees of freedom (and their
duals) of $D=11$ supergravity, which is $\cO_{E_8^{+++}}$.
\end{quote}

\subsection{Results}

The analysis above can be repeated for all very-extended algebras
$\lag^{+++}$. There is a natural maximal choice for the
$\mathfrak{sl}(D)$ algebra \cite{En,Kleinschmidt:2003mf}. This is
obtained by starting at the very-extended node and following the
line of long roots as far as possible.

The decomposition under this maximal gravity
subalgebra, truncated at the level of the affine root of
$\lag^+$, corresponds precisely to the bosonic fields of the
oxidized theory $\cO_G$, if one also includes {\em all} dual
fields for form matter and gravity \cite{Kleinschmidt:2003mf}.
Instead of repeating the full analysis we summarize the fields by
their oxidized theories in table \ref{fcres}; the Kac--Moody algebra
$\lag^{+++}$ captures only the bosonic fields. The truncation
criterion of the affine root seems natural since one knows that
the finite-dimensional algebras $\lag$ correspond to the oxidized
theories and generators in the `true Kac--Moody sector' are likely
to have a different r\^ole. \setcounter{table}{0}
\begin{table}
\centering
\begin{tabular}{|c|p{10cm}|}
\hline $\lag^{+++}$ & oxidized theory (maximal) \\
\hline
$E_8^{+++}$ & $N=1$, $D=11$ SUGRA\\
$E_7^{+++}$ & $N=IIB$, $D=10$ SUGRA, truncated non-supersymmetrically to
  dilaton and 4-form potential with self-dual field strength\\
$E_6^{+++}$ & $N=2$, $D=8$ SUGRA, truncated to dilaton,
  axion and 3-form potential\\
$D_n^{+++}$ & Massless sector of the closed bosonic string in $D=n+2$\\
$A_n^{+++}$ & Gravity in $D=n+3$\\
$G_2^{+++}$ & Einstein-Maxwell $N=2$, $D=5$ SUGRA\\
$F_4^{+++} $ & $N=(0,1)$, $D=6$ chiral SUGRA\\
$B_n^{+++} $ & massless `heterotic' string in $D=n+2$ $\equiv$
  closed bosonic string coupled to a massless abelian vector
  potential\\
$C_n^{+++} $ & $D=4$ theory, $(n-1)^2$ scalars and $2(n-1)$ vector
  potentials transforming under $C_n$\\
\hline
\end{tabular}
\caption{\label{fcres} The list of oxidized theories as recovered by
  very-extended Kac--Moody algebras $\lag^{+++}$.}
\end{table}

All oxidized theories are theories containing gravity and pure gravity
itself is associated with $A$ type algebras. It can be shown that on
the level of very-extended algebras the relevant maximal $A_{D-3}^{+++}$
algebra is contained in the M-theory algebra $\lag^{+++}$
\cite{Kleinschmidt:2003mf}. .

\subsection{Different embeddings}

We can analyse the field content of a theory with respect to
different embeddings, i.e. different choices of $A_n$
subalgebras. For example, the type IIA supergravity theory is
associated with the following decomposition of $E_8^{+++}$
(we now only mark in black the nodes to which we assign a level $\ell_i$),

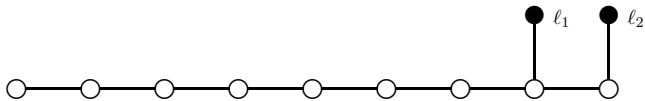
\begin{center}
\scalebox{.7}{
\begin{picture}(180,60)
  \put(300,45){$\ell_2$}
  \put(260,45){$\ell_1$}
\thicklines \multiput(-30,10)(40,0){9}{\circle{10}}
\multiput(-25,10)(40,0){8}{\line(1,0){30}}
\put(250,50){\circle*{10}} \put(250,15){\line(0,1){30}}
\put(290,50){\circle*{10}} \put(290,15){\line(0,1){30}}
\end{picture}
}
\end{center} The level $(\ell_1,\ell_2)$ generators of
$E_8^{+++}$ corresponding to the roots of the form
$\a  = \sum_s m_s \a_s + \ell_1 \a_{\ell_1} + \ell_2 \a_{\ell_2}$ are in
representations of $A_9$ characterized by the Dynkin labels
$\{p_s\} = [p_1,...,p_9]$. The following table gives, up to the
level of the affine root, the Dynkin labels of the
representations occurring in $E_8^{+++}$ and their outer
multiplicities $\mu$, as well as the interpretation of
the fields.

\begin{center}
\begin{tabular}{|c|c|r|l|}
\hline
$(\ell_1,\ell_2)$&$[p_1,...,p_9]$&$\mu$&Interpretation\\
\hline
(0,0)&[1,0,0,0,0,0,0,0,1]$\oplus$[0,0,0,0,0,0,0,0,0]&1&$h_a{}^b$\\
(0,0)&[0,0,0,0,0,0,0,0,0]&2&$\phi$  \\
(1,0)&[0,0,0,0,0,0,0,1,0]&1&$B_{(2)}$\\
(0,1)&[0,0,0,0,0,0,0,0,1]&1&$A_{(1)}$\\
(1,1)&[0,0,0,0,0,0,1,0,0]&1&$A_{(3)}$\\
(2,1)&[0,0,0,0,1,0,0,0,0]&1&$\tilde{A}_{(5)}$\\
(2,2)&[0,0,0,1,0,0,0,0,0]&1&$\tilde{B}_{(6)}$\\
(3,1)&[0,0,1,0,0,0,0,0,0]&1&$\tilde{A}_{(7)}$\\
(3,2)&[0,0,1,0,0,0,0,0,1]&1&$\tilde{A}_{(7,1)}$\\
(3,2)&[0,1,0,0,0,0,0,0,0]&1&$\tilde{A}^\phi_{(8)}$\\
... & & &\\
\hline
\end{tabular}
\end{center} These fields match the field content of
type IIA supergravity. Indeed, in addition to the level zero
gravitational fields and their dual fields, one recognizes the
NS-NS two form $B_{(2)}$, the RR one form $A_{(1)}$, the RR three
form $A_{(3)}$ and the dual $\tilde{B}_{(6)}$, $\tilde{A}_{(7)}$
and $\tilde{A}_{(5)}$, respectively.\newline

For type IIB supergravity, the choice of the $A_9$ subalgebra can
be depicted by redrawing the $E_8^{+++}$ Dynkin diagram in the
following way.\newline

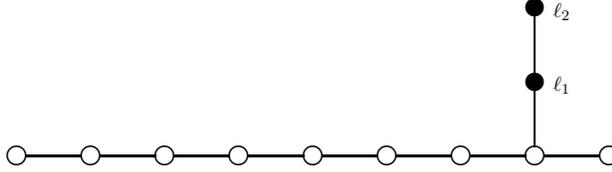
\begin{center}
\scalebox{.7}{
\begin{picture}(180,60)
  \put(260,85){$\ell_2$}
  \put(260,45){$\ell_1$}
\thicklines \multiput(-30,10)(40,0){9}{\circle{10}}
\multiput(-25,10)(40,0){8}{\line(1,0){30}}
\put(250,50){\circle*{10}} \put(250,15){\line(0,1){30}}
\put(250,90){\circle*{10}} \put(250,55){\line(0,1){30}}
\end{picture}
}
\end{center}
The decomposition produces the following table
\cite{Kleinschmidt:2003mf}.

\begin{center}
\begin{tabular}{|c|c|r|l|}
\hline
$(\ell_1,\ell_2)$&$[p_1,...,p_9]$&$\mu$&Interpretation\\
\hline
(0,0)&[1,0,0,0,0,0,0,0,1]$\oplus$[0,0,0,0,0,0,0,0,0]&1&$h_a{}^b$\\
(0,0)&[0,0,0,0,0,0,0,0,0]&2&$\phi$\\
(1,0)&[0,0,0,0,0,0,0,0,0]&1&$\chi$\\
(0,1)&[0,0,0,0,0,0,0,1,0]&1&$B_{(2)}$\\
(1,1)&[0,0,0,0,0,0,0,1,0]&1&$A_{(2)}$\\
(1,2)&[0,0,0,0,0,1,0,0,0]&1&$A_{(4)}$\\
(1,3)&[0,0,0,1,0,0,0,0,0]&1&$\tilde{B}_{(6)}$\\
(2,3)&[0,0,0,1,0,0,0,0,0]&1&$\tilde{A}_{(6)}$\\
(2,4)&[0,0,1,0,0,0,0,0,1]&1&$\tilde{A}_{(7,1)}$\\
(1,4)&[0,1,0,0,0,0,0,0,0]&1&$\tilde{A}^\chi_{(8)}$\\
(2,4)&[0,1,0,0,0,0,0,0,0]&1&$\tilde{A}^\phi_{(8)}$\\
... & & &\\
\hline
\end{tabular}
\end{center} The fields in the table correspond to the
gravitational fields  $h_a{}^b$ and the dilaton $\phi$,
the RR zero form (axion) $\chi$, the NS-NS two form $B_{(2)}$,
the RR two form $A_{(2)}$
and their duals  $\tilde{A}_{(7,1)}$, $\tilde{A}^\phi_{(8)}$,
$\tilde{A}^\chi_{(8)}$, $\tilde{B}_{(6)}$, $\tilde{A}_{(6)}$,
respectively. The RR four form $A_{(4)}$ will have self-dual  field
strength. Remember that the field content
corresponds only to the few lowest level generators. The tables
continue infinitely but there is no interpretation to date
for the additional fields, most of which have mixed Young symmetry
type.\footnote{We remark that there are two exceptions of fields which
  have a reasonable possible interpretation
  \cite{Kleinschmidt:2003mf}. These occur in the IIA and IIB
  decomposition of $E_8^{+++}$ and correspond to the nine-form
  potential in massive IIA supergravity and to the ten-form potential
  in IIB theory, which act as sources of the D8 in and D9 brane in IIA
  and IIB superstring theory, respectively.}
\newline

\noindent As string theories, type IIA and IIB string theories are
related by `T-duality'. Here we notice a diagrammatic reflection
of this fact through the `T-junction' in the Dynkin diagram of
$E_8^{+++}$. Therefore, the diagrams of the very-extended
Kac--Moody algebras nicely encode the field content of the
oxidized theories and also relations between different theories.

Moreover, Kaluza--Klein reduction corresponds to moving
nodes out of the `gravity line' (the $A$-type subalgebra). For
example, we saw that the bosonic field content of $D=11$
supergravity can be retrieved by considering the Dynkin diagram of
$E_8^{+++}$ with a $\mathfrak{sl}(11)$ subalgebra so that the
endpoint of the gravity line is the `M-theory' node in the diagram
below. A reduction of $D=11$ supergravity produces IIA
supergravity and the endpoint of the corresponding gravity line is
also marked in the diagram in agreement with the analysis above.
Finally, the ten-dimensional IIB theory is not a reduction of
$D=11$ supergravity but agrees with it after reduction to nine
dimensions. Roughly speaking this means taking a different
decompactification after $D=9$ not leading to M-theory. This is
precisely the structure of the $E_8^{+++}$ diagram.

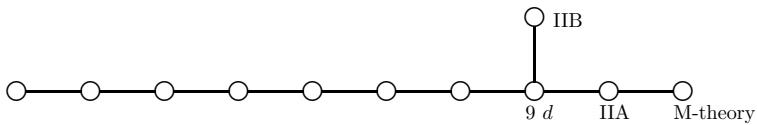
\begin{center}
\scalebox{.7}{
\begin{picture}(180,60)
  \put(245,-5){9 $d$}   \put(285,-5){IIA }
  \put(325,-5){M-theory}
  \put(260,45){IIB}
\thicklines \multiput(-30,10)(40,0){10}{\circle{10}}
\multiput(-25,10)(40,0){9}{\line(1,0){30}}
\put(250,50){\circle{10}} \put(250,15){\line(0,1){30}}
\end{picture}
}
\end{center}

\section{Relation to higher spin fields}

We now turn to the question in what sense the fields contained in the
Kac--Moody algebras are true higher spin fields. Our minimal criterion
will be that they transform under the {\em space-time} Lorentz group
and that their dynamics is consistent. These requirements will be
discussed for the two familiar classes of Kac--Moody models.\\

{\bf 1. West's proposal:} As explained in the introduction, in
\cite{West:2000ga,West:2001as} the Kac--Moody algebra is a symmetry of
the unreduced theory. Therefore, the fields transform under local
space-time $SO(D)\subset K(G^{+++})$ transformations and therefore
satisfy the
first requirement. But as we have also stressed, the derivation of the
dynamical equations requires the introduction of additional
translation generators and the closure with the conformal algebra to
obtain the correct curvatures, at least for gravity and the
anti-symmetric matter fields. It is not known what this procedure
yields for the mixed symmetry fields at higher levels. When writing
down the dynamical equations there is also an ambiguity in numerical
coefficients which should ultimately be fixed from the algebraic
structure alone, maybe together with supersymmetry.
At the present stage the consistency of the higher
spin dynamics cannot be determined.

However, it might be that known higher spin formulations play a r\^ole
in the resolution of these difficulties. In particular, the `unfolded
dynamics' of \cite{Vasiliev:2003ev,Vasiliev:2004qz}\footnote{See also
  M.~Vasiliev's contribution to the proceedings of this workshop.}
could provide a dynamical scheme for the KM fields.
\newline

{\bf 2. Null geodesic world-lines on Kac--Moody coset spaces:} This is the
approach taken in \cite{Damour:2002cu,Englert:2003py}, where
the idea is to map a one-dimensional world-line,
and not space-time itself, into the coset space $G^{+++}
/ K(G^{+++})$. The advantage of this approach is that one does not
require translation operators or the closure with the conformal
group. Rather space-time is conjectured to re-emerge
from a kind of Taylor expansion
via gradient fields of the oxidized fields present in the
decomposition tables. For
the details see \cite{Damour:2002cu} or the discussion in
\cite{Kleinschmidt:2003jf}. Therefore space-time is thought of as a
Kac--Moody intrinsic concept in the world-line approach.
The dynamical equations in this context are derived from a
lagrangian formulation for the particle motion.\\\indent
From the higher spin point-of-view the fields no longer are true
higher spin fields under the space-time Lorentz group but rather under
some internal Lorentz group. In order to transform the KM fields into
honest higher spin fields the recovery of space-time from the KM
algebra needs to be made precise. On the other hand, the dynamical
scheme here is already consistent and therefore our second requirement
for higher spin dynamics is fulfilled automatically.

\section{Space-time concepts}

Minkowski space-time can be seen as
the quotient of the Poincar\'e group by the Lorentz group. The
Poincar\'e group itself is the semi-direct product of the Lorentz
group with the (abelian) group of translations and the translations form a
vector representation of the Lorentz group:
\beq
[M^{ab},M^{cd}] &=&
    \eta^{ac}M^{bd}-\eta^{ad}M^{bc}
    +\eta^{bd}M^{ac}-\eta^{bc}M^{ad},\\{}
[P_a,P_b] &=& 0,\\{}
[M^{ab},P_c] &=& \d^a_c \eta^{bd}P_d -\d^b_c \eta^{ad} P_d.
\eeq

In the Kac-Moody context, the Lorentz group is replaced by
$K(G^{+++})$ so one needs a `vector representation' of
$K(G^{+++})$. By this we mean a representation graded as a vector
space by level
$\ell\ge 0$ with a Lorentz vector as bottom component at $\ell=0$,
corresponding to the vector space decomposition of the compact
subalgebra
\beq
K(\lag^{+++}) = \mathfrak{so}(D)\oplus ... \nonumber
\eeq
Unfortunately, very little is known about $K(\lag^{+++})$,
except that it is not a Kac--Moody algebra. However, it is evidently
a subalgebra of $\lag^{+++}$. One can therefore construct representations of
$K(\lag^{+++})$ by taking a representation of $\lag^{+++}$ and then
view it as a $K(\lag^{+++})$ module. Irreducibility (or complete
reducibility) of such  representations are interesting open
questions. If we take a (unitary) lowest weight representation of
$\lag^{+++}$ then we can write it as a tower of
$\mathfrak{gl}(D)$-modules in a fashion analogous to the decomposition
of the adjoint. The idea in reference
\cite{West:2003fc} was to take a representation of $\lag^{+++}$ whose
bottom component is a $\mathfrak{gl}(D)$ vector for any choice of
gravity subalgebra. This is naturally provided for by the
$\lag^{+++}$ representation  with lowest weight Dynkin labels
\beq \ [1,0,...,0]_{\lag^{+++}}. \no \eeq
We denote this irreducible $\lag^{+++}$
representation by $L(\lambda_1)$ since the
lowest weight is just the fundamental weight of the first
(very-extended) node.\footnote{Therefore, the representation is
  integrable and unitarizable.}
Decomposed w.r.t. the gravity subalgebra $A_{D-1}$, this yields
\beq  [\underbrace{1,0,\ldots,0}_{r \;
 \textrm{\footnotesize{labels}}}]_{\lag^{+++}} &\rightarrow&
 [\underbrace{1,0,\dots,0}_{(D-1) \;  \textrm{\footnotesize{labels}}}
 ]_{gravity} \oplus \ldots \no
\eeq where the "$\ldots$" denote the infinitely many other
representations of the gravity subalgebra at higher levels ($r$ is
the rank of $\lag^{+++}$). These fields can be computed in
principle with the help of a character formula. For example, for
$E_8^{+++}$, one finds the following decomposition of the
vectorial representation w.r.t. the M-theory gravity subalgebra
$A_{10}$ \cite{West:2003fc}:

\begin{center}
\begin{tabular}{|c|l|r|}
\hline
$\ell$&$[p_1,\ldots,p_{10}]$&Field\\
\hline
0 & $[1,0,0,0,0,0,0,0,0,0]$ & $P_a$ \\
1 & $[0,0,0,0,0,0,0,0,1,0]$ & $Z^{ab}$ \\
2 & $[0,0,0,0,0,1,0,0,0,0]$ & $Z^{a_1...a_5}$ \\
3 & $[0,0,0,1,0,0,0,0,0,1]$ & $Z^{a_1...a_7,b}$ \\
3 & $[0,1,0,0,0,0,0,0,0,0]$ & $Z^{a_1...a_8}$ \\
 & ... & \\
\hline\end{tabular}
\end{center} In addition to the desired vector representation
$P_a$ of $\mathfrak{gl}(11)$,
one finds a 2-form and a 5-form at levels $\ell=1,2$.
It was noted in \cite{West:2003fc}
that these are related to the $D=11$ superalgebra
\beq
\{ Q_\a,Q_\b \} = \big((\G^aC) P_a + \frac12 (\G_{ab}C) Z^{ab} +
\frac1{5!}(\G_{a_1\ldots a_5}C) Z^{a_1\ldots a_5}\big)_{\a\b}
\eeq
in an obvious way. ($C$ is the charge conjugation matrix.)
The branes of M-theory couple as $\frac12$-BPS
states to these central charges.
Therefore the vector representation also encodes information about the
topological charges of the oxidized theory. We also note that the
charges on $\ell=3$ suggest that they might be carried by solitonic
solutions associated with the dual graviton field, but no such
solutions are known.\footnote{This idea has appeared before in the
  context of U-duality where the mixed symmetry `charge' was
  associated with a Taub-NUT solution \cite{Obers:1998fb}.} Therefore,
we will refer to the tensors in the $\lag^{+++}$ representation
$L(\l_1)$ as `generalized
central charges' \cite{Kleinschmidt:2003jf}.\footnote{These are
  generalized since it is not clear in what algebra they should be
  central and the name is just by analogy. Note, however, that these
  charges will still all commute among themselves.}\newline

The result carries over to all algebras $\lag^{+++}$: To any field we
found in the decomposition for the oxidized theory there is an
associated generalized charge \cite{Kleinschmidt:2003jf} in the
`momentum representation'' $L(\lambda_1)$.
For instance, for an anti-symmetric $p$-form this is just a
$(p-1)$-form. It is remarkable that this result holds for all
$\lag^{+++}$ and not only for those where one has a supersymmetric
extension of the oxidized theory.
\newline

To demonstrate this result we use a trick \cite{Kleinschmidt:2003jf}
which turns the statement
into a simple corollary of the field content analysis. Instead of using
the Weyl--Kac or the Freudenthal character formulae, we embed the
semi-direct sum of the algebra with its vectorial representation
in a larger algebra \cite{Kleinschmidt:2003jf}. This can be
illustrated nicely for the Poincar\'e algebra. The aim is to find a
minimal algebra which contains (as a truncation) both the original
Lorentz  algebra and the vector representation by Dynkin diagram
extensions. We  consider the case of even space-time dimension for
simplicity. The  $\mathfrak{so}(2d)$ vector has Dynkin labels
$[1,0,\ldots,0]$ in the  standard choice of labelling the simple
nodes. Hence, adding a node (marked in black in the diagram below,
with label $m$ for momentum) at
the position $1$ below with  a single line will give a vector in the
decomposition under the Lorentz algebra at $\ell=1$.

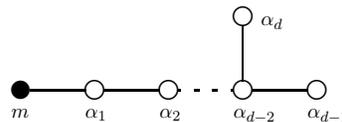
\begin{center}
\scalebox{.7}{
\begin{picture}(180,60)
\put(5,-5){$m$} \put(45,-5){$\alpha_1$}
 \put(125,-5){$\alpha_{d-2}$}  \put(85,-5){$\alpha_{2}$}
  \put(140,45){$\alpha_d$}\put(165,-5){$\alpha_{d-1}$}
\put(5,-5){} \thicklines \put(10,10){\circle*{10}}
 \multiput(15,10)(40,0){2}{\line(1,0){30}}
\dashline[0]{2}(95,10)(105,10)(115,10)(125,10)
\put(130,50){\circle{10}} \put(130,15){\line(0,1){30}}

\multiput(50,10)(40,0){4}{\circle{10}}
\put(135,10){\line(1,0){30}}
\end{picture}
} \end{center}
The resulting embedding algebra is the conformal algebra as the full
decomposition shows\footnote{The diagram
  only shows this for the complex algebras but one can actually check
  it at the level of real forms as well.}
\beq \{ M^{ab}, P^c \} \rightarrow_{embed} M^{AB}
\rightarrow_{decompose} \{ M^{ab}, P^c, K^c, D \},
\eeq
where  $M^{ab}$ ($a=1,...,2d$), $P^a$, $K^a$ and $D$  are respectively
the  generators of the Lorentz group, the translations generators, the
special conformal transformations generators and the dilatation
operator in $D$ dimensions. $M^{AB}$ ($A=1,...,2d+2$) are the
generators of the conformal algebra $\mathfrak{so}(2d+1,1)$. (Note
that if one allows a change in the original diagram  then there is
smaller embedding in the AdS algebra $\mathfrak{so}(2d,1)$
\cite{MacDowell:1977jt}; a fact that has also been exploited in the
higher spin literature.)\newline

\noindent Returning to the KM case, one embeds the semi-direct product
of $\lag^{+++}$ with its vector representation in a larger algebra in
a similar fashion. The resulting algebra is obtained by adding one
more node to the Dynkin diagram of $\lag^{+++}$ at the very-extended
node and we denote this fourth extension of $\lag$ by $\lag^{++++}$:

\beq
\lag^{+++} \ltimes L(\lambda_1) \hookrightarrow \lag^{++++}.
\eeq
Note that the embedding is only valid at the level of vector spaces;
the Lie algebra relations in the two spaces are different, since not
all the central charges will commute any longer in
$\lag^{+++}$. (However, the momenta $P_a$ will commute.)
The Dynkin diagram of $\lag^{++++}$, for the very-extension
$E^{+++}_8$, is the following:

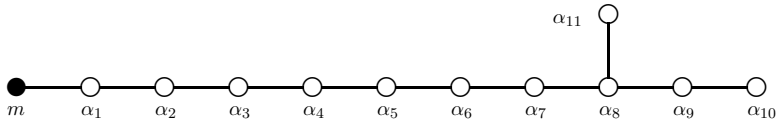
\scalebox{.7}{
\begin{picture}(180,60)
\put(-5,-5){$m$}
\put(35,-5){$\alpha_{1}$} \put(75,-5){$\alpha_2$}
\put(115,-5){$\alpha_3$}
 \put(155,-5){$\alpha_4$}
  \put(195,-5){$\alpha_5$} \put(235,-5){$\alpha_6$}
  \put(275,-5){$\alpha_7$}   \put(315,-5){$\alpha_8$}
  \put(355,-5){$\alpha_9$} \put(395,-5){$\alpha_{10}$}
  \put(290,45){$\alpha_{11}$}
\thicklines \multiput(40,10)(40,0){10}{\circle{10}}
\multiput(5,10)(40,0){10}{\line(1,0){30}}
\put(320,50){\circle{10}} \put(320,15){\line(0,1){30}}
\put(0,10){\circle*{10}}
\end{picture}
}\newline

\noindent By using this kind of embedding, it follows that one
obtains all generalized central charges for the KM fields of the
oxidized theories. This is a consequence of a Kaluza--Klein
reduction of the oxidized fields: A field in $D$ space-time
dimensions gets lifted to one in $D+1$ dimensions by the diagram
extension and subsequent reduction then results in the field in
$D$ dimensions and its corresponding charge. We remark that in
this construction of space-time there are infinitely many
space-time generators, rendering space-time infinite-dimensional.
(The idea of associating space-time directions to central charges
goes back to \cite{So78}.)

\section{Conclusions}

We have demonstrated that indefinite Kac--Moody algebras
are a rich natural source for higher spin
fields with mixed symmetry type. In all existing proposals, however,
there are important open problems in deriving true and consistent
higher spin field theories from this algebraic approach to M-theory.
As we indicated, known results from higher spin theory might be used to
solve some of these problems. Further progress might be derived from
an understanding of the KM structure to all levels which appears to be
a very hard problem.

All the models we discussed so far are for bosonic fields, an
extension to fermionic degrees of freedom relies on an understanding
of the compact subalgebra $K(\lag)$. The next step then would be to
relate these to the fermionic (or supersymmetric)
HS theories {\em \`a la} Fang--Fronsdal.\\

{\bf Acknowledgements:} SdB is grateful to Chistiane Schombond for
her careful reading of the manuscript and her precious help. AK is
grateful to Peter West and Igor Schnakenburg
for numerous discussions and enjoyable collaborations. AK
would like to thank the organizers for the invitation to present
these results and for providing an interesting and stimulating
workshop. The work of AK was supported in part by the
`Studienstiftung des deutschen Volkes'. The work of SdB is
supported in part by the ``Actions de Recherche Concert{\'e}es" of
the ``Direction de la Recherche Scientifique - Communaut{\'e}
Fran{\c c}aise de Belgique", by a ``P\^ole d'Attraction
Interuniversitaire" (Belgium), by IISN-Belgium (convention
4.4505.86). Support from the European Commission RTN programme
HPRN-CT-00131, in which SdB is associated to K. U. Leuven, is also
acknowledged.

\appendix
\section{Coset Models}

Let us recall some basic facts about non-linear sigma models
associated with the coset model $G / K(G)$. We use the Cartan
decomposition of the Lie algebra $\lag$ of $G$,

$$ \lag = \lap+\lak $$

\noindent where $\lak$ is the Lie algebra corresponding to
$K(G)$ and is fixed by an involution $\o$ of $\lag$:

$$ \lak = \{ x \in \lag \ \arrowvert \ \o(x) = x \}, $$
$$ \lap = \{x \in \lag \ \arrowvert \ \o(x) = -x \}. $$

\noindent We will here take $\o$ to be the compact involution,
i.e. a map from $\lag$ to $\lag$ such that

$$ \o(e_i) = -f_i \ ; \ \o(f_i) = -e_i \ ; \ \o(h_i)=-h_i$$
on the $r$ Chevalley generators of $\lag$. Let us consider the coset
element $v$ in Iwasawa parametrization

$$ v(x) = \exp(\overrightarrow{\phi}(x) . \overrightarrow{h} ) \exp{(
\sum_{\alpha \in \Delta_+} A_{\alpha}(x) E^{\alpha}}) \in G/K(G)
$$

\noindent where the $\overrightarrow{h}$ are the generators of the
Cartan subalgebra and the $E^{\alpha}$'s are the step operators
associated with the positive roots. The coordinates on $G/K(G)$
namely $\overrightarrow{\phi}$ and $A_{\alpha}$ depend on the
space-time coordinates. $dv v^{-1}$ is in $\lag$, and hence
we can decompose it as $dv v^{-1} = \mathcal{P} + \mathcal{Q} =
(\mathcal{P}_{\m} + \mathcal{Q}_{\m}) dx^{\m}$. A natural
lagrangian associated to this model is

$$ L = {1\over 4n} (\mathcal{P}_{\mu} \arrowvert
\mathcal{P}^{\mu})$$ where the inner product $( \cdot|\cdot )$ is the
invariant metric on $\lag$. Such a lagrangian is invariant under
local $K(G)$ transformations and global $G$ transformations,

$$ v(x) \ \rightarrow \ k(x) v(x) g$$

\noindent where $k(x)$ is in $K(G)$ and $g$ is in $G$. The
lagrangian equations of motion are
\setcounter{equation}{0}
\beq \label{eom}
D^{\mu}(n^{-1} \mathcal{P}_{\mu}) = ( \partial^{\mu} - \textrm{ad}\,
\mathcal{Q}^{\mu} ) (n^{-1} \mathcal{P}_{\mu}) = 0 \eeq
Furthermore, the Lagrange multiplier $n$ constrains the motion to be
null, which is covariantly constant along the coset by (\ref{eom}).
An introduction to scalar coset models can be found in \cite{Pope}.


\begin{thebibliography}{999}


\bibitem{Gross:1988ue} D.~J.~Gross, {\sl High-Energy Symmetries Of String
  Theory}, Phys.\ Rev.\ Lett.\  {\bf 60} (1988) 1229.

\bibitem{Cremmer:1997ct} E.~Cremmer, B.~Julia, H.~Lu and C.~N.~Pope,
  {\sl Dualisation of dualities. I}, Nucl.\ Phys.\ B {\bf 523}
  (1998) 73 [arXiv:hep-th/9710119]. E.~Cremmer, B.~Julia, H.~Lu and C.~N.~Pope,
  {\sl Dualisation of dualities. II: Twisted self-duality of doubled
  fields  and superdualities}, Nucl.\ Phys.\ B {\bf 535} (1998) 242
  [arXiv:hep-th/9806106].

\bibitem{West:2001as} P.~C.~West, {\sl E(11) and M theory}, Class.\
  Quant.\ Grav.\  {\bf 18} (2001) 4443 [arXiv:hep-th/0104081].

\bibitem{BJall1} P.~Henry-Labordere, B.~Julia and L.~Paulot, {\sl
  Borcherds symmetries in M-theory},  JHEP {\bf 0204} (2002) 049
  [arXiv:hep-th/0203070].

\bibitem{Damour:2002cu} T.~Damour, M.~Henneaux and H.~Nicolai, {\sl
  E(10) and a 'small tension expansion' of M theory}, Phys.\ Rev.\
  Lett.\  {\bf 89} (2002) 221601 [arXiv:hep-th/0207267].

\bibitem{Englert:2003py} F.~Englert and L.~Houart, {\sl G+++ invariant
  formulation of gravity and M-theories: Exact BPS solutions}, JHEP
  {\bf 0401} (2004) 002 [arXiv:hep-th/0311255].

\bibitem{Ju82} B. Julia, {\sl Kac-Moody Symmetry Of Gravitation And
  Supergravity Theories}, invited talk given at AMS-SIAM Summer Seminar
  on Applications of Group Theory in Physics and Mathematical Physics,
  Chicago, Ill., Jul 6--16, 1982; in: Vertex Operators in Mathematics
  and Physics, Publications of the Mathematical Sciences Research
  Institute no 3, Springer Verlag (1984);

\bibitem{Hull:1994ys} C.~M.~Hull and P.~K.~Townsend, {\sl Unity of
  superstring dualities}, Nucl.\ Phys.\ B {\bf 438} (1995) 109
  [arXiv:hep-th/9410167].

\bibitem{BrMa87} P.~Breitenlohner and D.~Maison, {\sl On the
    Geroch group}, Ann. Poincar\'e Phys. Theor. {\bf 46} (1987)
    215--246. H.~Nicolai, {\sl The integrability of $N=16$ supergravity},
    Phys. Lett. B {\bf 194} (1987) 402--407.

\bibitem{Ju81a} B.~Julia, {\sl Group Disintegrations}, in:
  S.~W.~Hawking and M.~Ro\v{c}ek (eds.), Superspace and
  Supergravity, Proceedings of the Nuffield Workshop,
  Cambridge, Eng., Jun 22 -- Jul 12, 1980, Cambridge
  University Press (Cambridge, 1981) 331--350, LPTENS
   80/16

\bibitem{Cremmer:1999du} E. Cremmer, B. Julia, H. Lu and C.N. Pope,
 {\sl Higher-dimensional origin of D=3 coset symmetries},
 [arXiv:hep-th/9909099].

\bibitem{Ke03a} A.~Keurentjes, {\sl Group Theory of Oxidation},
  Nucl. Phys. B {\bf 658} (2003) 303--347, [arXiv:hep-th/0210178];
  {\sl Group Theory of Oxidation 2:
  Cosets of Nonsplit Groups}, Nucl. Phys. B {\bf 658} (2003)
  348--372, [arXiv:hep-th/0212024]

\bibitem{Belinsky:1970ew} V.~A.~Belinsky, I.~M.~Khalatnikov and
E.~M.~Lifshitz, {\sl Oscillatory Approach To A Singular Point In
The Relativistic Cosmology}, Adv.\ Phys.\  {\bf 19} (1970) 525.

\bibitem{Belinsky:1982pk} V.A.~Belinsky, I.M.~Khalatnikov and
E.M.~Lifshitz, {\sl A general solution of the Einstein equations
with a time  singularity}, Adv.\ Phys.\ {\bf 31} (1982) 639.

\bibitem{Damour:2002et} T.~Damour, M.~Henneaux and H.~Nicolai,
{\sl Cosmological billiards}, Class.\ Quant.\ Grav.\  {\bf 20}
(2003) R145 [arXiv:hep-th/0212256].

\bibitem{Damour:2002fz} T.~Damour, S.~de Buyl, M.~Henneaux and
  C.~Schomblond, {\sl Einstein billiards and overextensions of
  finite-dimensional simple Lie algebras}, JHEP {\bf 0208} (2002) 030
  [arXiv:hep-th/0206125].

\bibitem{Damour:2001sa} T.~Damour, M.~Henneaux, B.~Julia and
H.~Nicolai, {\sl Hyperbolic Kac-Moody algebras and chaos in
Kaluza-Klein models}, Phys.\ Lett.\ B {\bf 509} (2001) 323
[arXiv:hep-th/0103094].

\bibitem{Gaberdiel:2002db} M.~R.~Gaberdiel, D.~I.~Olive and P.~C.~West,
{\sl A class of Lorentzian Kac-Moody algebras}, Nucl.\ Phys.\ B {\bf
645} (2002) 403 [arXiv:hep-th/0205068].

\bibitem{En} F.~Englert, L.~Houart, A.~Taormina and P.~West, {\sl
The Symmetry of M-Theories}, JHEP {\bf 0309} (2003) 020
[arXiv:hep-th/0304206].

\bibitem{Kleinschmidt:2003mf} A.~Kleinschmidt, I.~Schnakenburg
and P.~West, {\sl Very-extended Kac-Moody algebras and their
interpretation at low levels}, Class.\ Quant.\ Grav.\
{\bf 21} (2004) 2493 [arXiv:hep-th/0309198].

\bibitem{West:2000ga} P.~C.~West, {\sl Hidden superconformal
symmetry in M theory}, JHEP {\bf 0008} (2000) 007
[arXiv:hep-th/0005270].

\bibitem{Og73} V.~I.~Ogievetsky, {\sl Infinite-dimensional algebra
    of general covariance group as the closure of
    finite-dimensional algebras of conformal and linear groups},
    Lett. Nuovo Cim. {\bf 8} (1973) 988--990.

\bibitem{SchnWe01} I.~Schnakenburg and P.~C.~West, {\sl Kac--Moody
    symmetries of 2B supergravity}, Phys. Lett. B {\bf 517} (2001)
    421--428 [arXiv:hep-th/0107081].

\bibitem{SchnWe02} I.~Schnakenburg and P.~C.~West, {\sl Massive
    IIA supergravity as a nonlinear realization}, Phys. Lett. B
    {\bf 540} (2002) 137--145 [arXiv:hep-th/0204207].

\bibitem{LaWe01} N.~D.~Lambert and P.~C.~West, {\sl Coset
    symmetries in dimensionally reduced bosonic string theory},
    Nucl. Phys. B {\bf 615} (2001) 117--132 [arXiv:hep-th/0107029].

\bibitem{EnHoWe03} F.~Englert, L.~Houart and P.~West, {\sl
    Intersection rules, dynamics and symmetries}, JHEP {\bf 0308}
    (2003) 025 [arXiv:hep-th/0307024].

\bibitem{West:2004st} P.~West, {\sl `The IIA, IIB and eleven
  dimensional theories and their common E(11) origin}, Nucl.\ Phys.\ B
  {\bf 693} (2004) 76 [arXiv:hep-th/0402140].

\bibitem{West:2004kb} P.~West, {\sl E(11) origin of brane charges and
  U-duality multiplets}, JHEP {\bf 0408} (2004) 052 [arXiv:hep-th/0406150].

\bibitem{Kac} V.Kac, {\sl Infinite Dimensional Lie Algebras},
third edition, Cambridge University Press (1990).

\bibitem{NiFi03} H.~Nicolai and T.~Fischbacher, {\sl Low level
    representations of $E_{10}$ and $E_{11}$}, Contribution to the
    Proceedings of the Ramanujan International Symposium on
    Kac--Moody Algebras and Applications, ISKMAA-2002, Chennai,
    India, 28--31 January [arXiv:hep-th/0301017].

\bibitem{Keurentjes:2004bv} A.~Keurentjes, {\sl E(11): Sign of the
times}, [arXiv:hep-th/0402090].

\bibitem{Keurentjes:2004xx} A.~Keurentjes, {\sl Time-like T-duality
algebra}, [arXiv:hep-th/0404174].

\bibitem{KlNi04} A.~Kleinschmidt and H.~Nicolai, {\sl $E_{10}$ and
    $SO(9,9)$ invariant supergravity}, JHEP {\bf 0407} (2004) 041
    [arXiv:hep-th/0407101].

\bibitem{Hull:2001iu} C.~M.~Hull, {\sl Duality in gravity and
higher spin gauge fields}, JHEP {\bf 0109} (2001) 027
[arXiv:hep-th/0107149].

\bibitem{Curtright:1980yk} T.~Curtright, {\sl Generalized Gauge
Fields},  Phys.\ Lett.\ B {\bf 165} (1985) 304.

\bibitem{Bekaert:2002uh} X.~Bekaert, N.~Boulanger and M.~Henneaux,
{\sl Consistent deformations of dual formulations of linearized
gravity: A no-go result}, Phys.\ Rev.\ D {\bf 67} (2003) 044010
[arXiv:hep-th/0210278].

\bibitem{Vasiliev:2003ev} M.~A.~Vasiliev, {\sl Nonlinear equations for
  symmetric massless higher spin fields in  (A)dS(d)},  Phys.\ Lett.\
  B {\bf 567} (2003) 139 [arXiv:hep-th/0304049].

\bibitem{Vasiliev:2004qz} M.~A.~Vasiliev, {\sl Higher spin gauge
  theories in various dimensions}, Fortsch.\ Phys.\  {\bf 52} (2004)
  702 [arXiv:hep-th/0401177].

\bibitem{West:2003fc} P.~West, {\sl E(11), SL(32) and central charges},
 Phys.\ Lett.\ B {\bf 575} (2003) 333 [arXiv:hep-th/0307098].

\bibitem{Kleinschmidt:2003jf} A.~Kleinschmidt and P.~West,
{\sl Representations of G+++ and the role of space-time},
JHEP {\bf 0402} (2004) 033 [arXiv:hep-th/0312247].

\bibitem{MacDowell:1977jt} S.~W.~MacDowell and F.~Mansouri,
{\sl Unified Geometric Theory Of Gravity And Supergravity},
Phys.\ Rev.\ Lett.\  {\bf 38} (1977) 739 [Erratum-ibid.\  {\bf 38}
(1977) 1376].

\bibitem{Obers:1998fb} N.~A.~Obers and B.~Pioline, {\sl U-duality and
  M-theory},  Phys.\ Rept.\  {\bf 318} (1999) 113 [arXiv:hep-th/9809039].

\bibitem{So78} M.~F.~Sohnius, {\sl Supersymmetry and central
  charges}, Nucl. Phys. B {\bf 138} (1978) 109--121

\bibitem{Pope} C.~N.~Pope, {\sl Kaluza--Klein Theory}, Lecture notes available
online under {\tt
http://faculty.physics.tamu.edu/\~{}pope/ihplec.ps}

\end{thebibliography}
\end{document}